\documentclass[twocolumn,a4paper,amsmath,amssymb]{revtex4}
\usepackage[urlcolor=blue, hyperindex, colorlinks, bookmarks=true]{hyperref}
\usepackage{graphicx}
\usepackage{times}
\usepackage{amsmath}
\usepackage{braket}
\usepackage{xcolor}
\usepackage{tabularx}
\usepackage[symbol]{footmisc}
\renewcommand{\eqref}[1]{Eq.~(\ref{#1})}

\newcommand{\figref}[1]{Fig.~\ref{#1}}

\newcommand{\removedD}[1]{{\color{gray}{#1}}}
\renewcommand{\removedD}[1]{{}} 

\newcommand{\corrected}[1]{}

\renewcommand{\eqref}[1]{Eq.~(\ref{#1})}

\newcommand{\appref}[1]{\hyperref[#1]{Appendix~\ref*{#1}}}
\newcommand{\tabref}[1]{\hyperref[#1]{Table~\ref*{#1}}}

\usepackage[english]{babel}
\begin{document}
\title{Realizing a Circuit Analog of an Optomechanical System\\
with Longitudinally Coupled  Superconducting Resonators}
\author{C. Eichler$^1$\footnote{Present address: eichlerc@phys.ethz.ch, ETH Zurich} and J. R. Petta$^1$}
\affiliation{$^1$Department of Physics, Princeton University, Princeton, New Jersey, 08544}
%
\date{\today}
\begin{abstract}
{
We realize a superconducting circuit analog of the generic cavity-optomechanical Hamiltonian by longitudinally coupling two superconducting resonators, which are an order of magnitude different in frequency. We achieve longitudinal coupling by embedding a superconducting quantum interference device (SQUID) into a high frequency resonator, making its resonance frequency depend on the zero point current fluctuations of a nearby low frequency $LC$-resonator. By employing sideband drive fields we enhance the intrinsic coupling strength of about 15$\,$kHz up to 280$\,$kHz by controlling the amplitude of the drive field. Our results pave the way towards the exploration of optomechanical effects in a fully superconducting platform and could enable quantum optics experiments with photons in the yet unexplored radio frequency band.
}
\end{abstract}
\maketitle
Circuit quantum electrodynamics (cQED) has become one of the primary platforms used to experimentally explore fundamental aspects of quantum physics \cite{Ansmann2009,Hofheinz2009,Astafiev2010,Vijay2011}, build practical devices for sensitive measurements \cite{Castellanos2008,Bergeal2010,Hatridge2011,Eichler2014,Macklin2015}, and eventually realize fault tolerant quantum computers \cite{Devoret2013,DiCarlo2009,Reed2012,Barends2014,Ofek2016}. The versatility in the design and fabrication of these circuits has also enabled their efficient coupling to other quantized degrees of freedom such as spins and charges in semiconductors \cite{Kubo2010,Schuster2010,Frey2012,Petersson2012a,Bienfait2016a,Eichler2017,Mi2017} and mechanical resonators \cite{Teufel2011,Palomaki2013,Pirkkalainen2013,Toth2017}, as well as their use for the sensing of electromagnetic noise \cite{Astafiev2004a,Yoshihara2006,Bylander2011,Bal2012}.

Individual elements in cQED devices, such as resonators and qubits, are most commonly coupled to each other through \emph{field-field} or \emph{dipole-field} interactions,  which typically result in Jaynes-Cummings-type coupling Hamiltonians of the form $H_{\rm int} \sim a^\dagger b + a b^\dagger$, where $a$ ($b$) and $a^\dagger$ ($b^\dagger$) are annihilation and creation operators of the two coupled modes, respectively. Such couplings are also referred to as {transversal} couplings \cite{Richer2016}, relating the orientation of the qubit dipole operator to the quantization axis defined by the uncoupled qubit eigenstates. At large detunings between two transversally coupled systems the presence of strong anharmonicities gives rise to effective \emph{energy-energy} interactions of the form $H_{\rm int} \sim a^\dagger a b^\dagger b$, also known as dispersive \cite{Blais2004,Schuster2007a} or cross-Kerr \cite{Kumar2009,Holland2015} interactions. Such interactions play a crucial role in the state-of-the-art dispersive readout of qubits \cite{Walter2017} and in the cat state paradigm of quantum computing \cite{Vlastakis2013}.

\begin{figure}[b]
\includegraphics[scale=0.9]{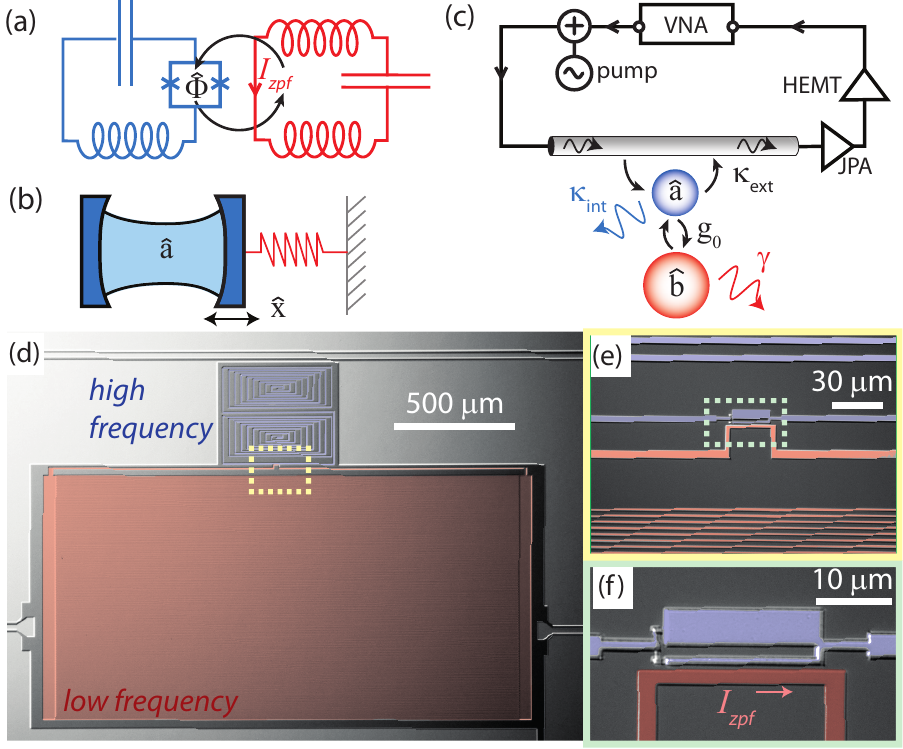}
\caption{ (a) Circuit representation of the device architecture.
(b) Analogous schematic of a generic optomechanical system. One of the cavity mirrors is movable and attached to a spring (red). A displacement of the mirror $x$ thereby shifts the resonance frequency of the cavity leading to an effective coupling between the radiation field and the motion of the mechanical oscillator. (c) Schematic of the experimental setup.
Resonator $a$ is side-coupled to a feedline with rate $\kappa_{\rm ext}$ used for driving and probing the system with a vector network analyzer (VNA). (d)-(f) Optical images of the sample at three different magnifications showing both resonators in (d), and the coupling region in (e) and (f). Two additional gate lines coupled to the low frequency resonator have been used to directly probe the low frequency resonance of similar devices at $T=1.8\,$K.}
\label{fig:1}
\end{figure}

As a complement to these established coupling schemes,  it has recently been proposed to couple the \emph{field} of one mode to the \emph{energy} of another \cite{Johansson2014a} -- a coupling mechanism often referred to as longitudinal coupling, which has the form $H_{\rm int} \sim (a+a^\dagger) b^\dagger b$. Such longitudinal coupling could prove useful in the readout of superconducting qubits \cite{Didier2015c}, and for the realization of fast and scalable two-qubit gates \cite{Royer2017,Richer2016,Chow2011}. While this type of coupling is unconventional for superconducting circuit systems and has thus far mostly been studied theoretically, it constitutes the archetypical interaction mechanism in cavity-optomechanical systems \cite{Aspelmeyer2013}. Here, the interaction stems from a frequency shift of the cavity which is induced by the displacement field of a mechanical oscillator. Realizing an analogous coupling scheme with superconducting circuits could enable a variety of optomechanically inspired experiments, ranging from ground state cooling of low frequency electromagnetic modes \cite{Teufel2011}, to coherent frequency conversion and amplification \cite{Lecocq2016}, and the development of nonreciprocal devices such as directional amplifiers and circulators \cite{Barzanjeh2017}.

In this letter, we demonstrate longitudinal coupling between a low frequency $LC$-resonator at $\omega_b/2\pi\approx 584\,$MHz, and a frequency tunable resonator around $\omega_a/2\pi\approx 5.4\,$GHz. The interaction arises from the zero-point fluctuations of the current in the low frequency resonator causing a change of the magnetic flux in the SQUID loop embedded in the high frequency resonator \cite{Johansson2014a} [see Fig.~\ref{fig:1}(a)]. The same coupling mechanism has previously been explored in a classical regime for magnetometry \cite{Hatridge2011} and to pump parametric amplifiers \cite{Yamamoto2008,Zhong2013}. Here, we demonstrate experiments in the quantum regime, where the magnetic flux in the SQUID loop arises from a quantum rather than a classical field.

The basic coupling mechanism is illustrated in the circuit schematic shown in \figref{fig:1}(a). The current flowing through the inductive wire of the low frequency resonator (red) generates a magnetic flux $\Phi$ that couples into the SQUID loop embedded in the high frequency resonator (blue). When the high frequency resonator is biased close to half a flux quantum $\Phi_{\rm ext}\approx \Phi_0/2 \equiv h/(4e)$, the frequency sensitively depends on this additional quantized flux, mediating a coupling described by the Hamiltonian
\begin{equation}
{H}_{\rm int}/\hbar = \Phi_{\rm zpf} \frac{\partial \omega_{a}}{\partial \Phi_{\rm ext}} a^\dagger a (b + b^\dagger).
\label{eq:Hint}
\end{equation}
Here,  $a$ ($b$) and $a^\dagger$ ($b^\dagger$) are the annihilation and creation operators of the high (low) frequency resonator, respectively,  $\omega_{a}$ is the frequency of the high frequency resonator, and $\Phi_{\rm zpf}$ is the portion of magnetic flux threading the SQUID loop associated with the zero point fluctuations of the low frequency resonator. The bare coupling strength $g_0 = \Phi_{\rm zpf} \frac{\partial \omega_{a}}{\partial \Phi_{\rm ext}}$ is thus given by the product of $\Phi_{\rm zpf}$ and the sensitivity of the resonance frequency to flux. In the above expression we have already taken into account that $\Phi_{\rm zpf}\ll\Phi_0$, and we can thus linearize the flux dependent resonance frequency around the bias point $\Phi_{\rm ext}$. The Hamiltonian~(\ref{eq:Hint}) is of the same form as the generic cavity optomechanical interaction Hamiltonian \cite{Aspelmeyer2013}. In this case, the frequency shift of a cavity is caused by the displacement of a mechanical oscillator as schematically shown in \figref{fig:1}(b). The similarity of these two systems may enable the exploration of quantum optics experiments with resonators at radio frequencies, which could for example be useful to couple to other degrees of freedom with a low transition frequency, such as nuclear spins in a Zeeman field.

The sample used in our experiments consists of two resonators, one of which is side-coupled to a coplanar waveguide used for driving and probing the system [see \figref{fig:1}(c)-(f)]. The high frequency resonator is formed by two spiral inductors with a SQUID in the middle, which are all fabricated using electron beam lithography and shadow evaporated aluminum. The low frequency resonator is formed by a large interdigitated finger capacitor with a simulated capacitance of  $C \approx 40\,$pF in parallel with a $2\,$mm long and $2\,\mu$m wide inductive wire that passes the SQUID at a distance of $d\approx3\,\mu{\rm m}$. The low frequency resonator and all other elements on the sample are fabricated from a sputtered niobium thin film using electron beam lithography and reactive ion etching. The zero point fluctuations of the current flowing through the inductive wire $I_{\rm zpf} = \sqrt{\hbar \omega_{b}^3 C/2}$ generate a magnetic field of approximately $B_{\rm zpf} = \mu_0 I_{\rm zpf}/(2\pi d)$ at a distance $d$, which in turn causes a magnetic flux in the SQUID loop $\Phi_{\rm zpf} = A B_{\rm zpf}$ proportional to the SQUID loop area $A$. For our sample parameters $\omega_{b}/2\pi\approx 584\,$MHz,  $A\approx 27 \mu$m$^2$, $d\approx 3 \,\mu$m, we find an approximate value of $\Phi_{\rm zpf} \approx 9\mu\Phi_0$.

The sample is mounted at the base plate of a dilution refrigerator cooled down to $T=20\,$mK and protected from external magnetic noise using two layers of cryoperm shielding. The input and output of the sample are connected to a standard microwave frequency measurement setup after several stages of amplification and probed either with a vector network analyzer (VNA) or using analog-to-digital conversion and field programmable gate array (FPGA) based electronics. Additional pump fields  from a microwave signal generator are applied through the same input line using a power combiner at room temperature. The input signal is strongly attenuated with a chain of cold attenuators and the output signal passes through a chain of low noise amplifiers including a Josephson parametric amplifier \cite{Eichler2014a}.
\begin{figure}[b!]
\includegraphics[scale=0.92]{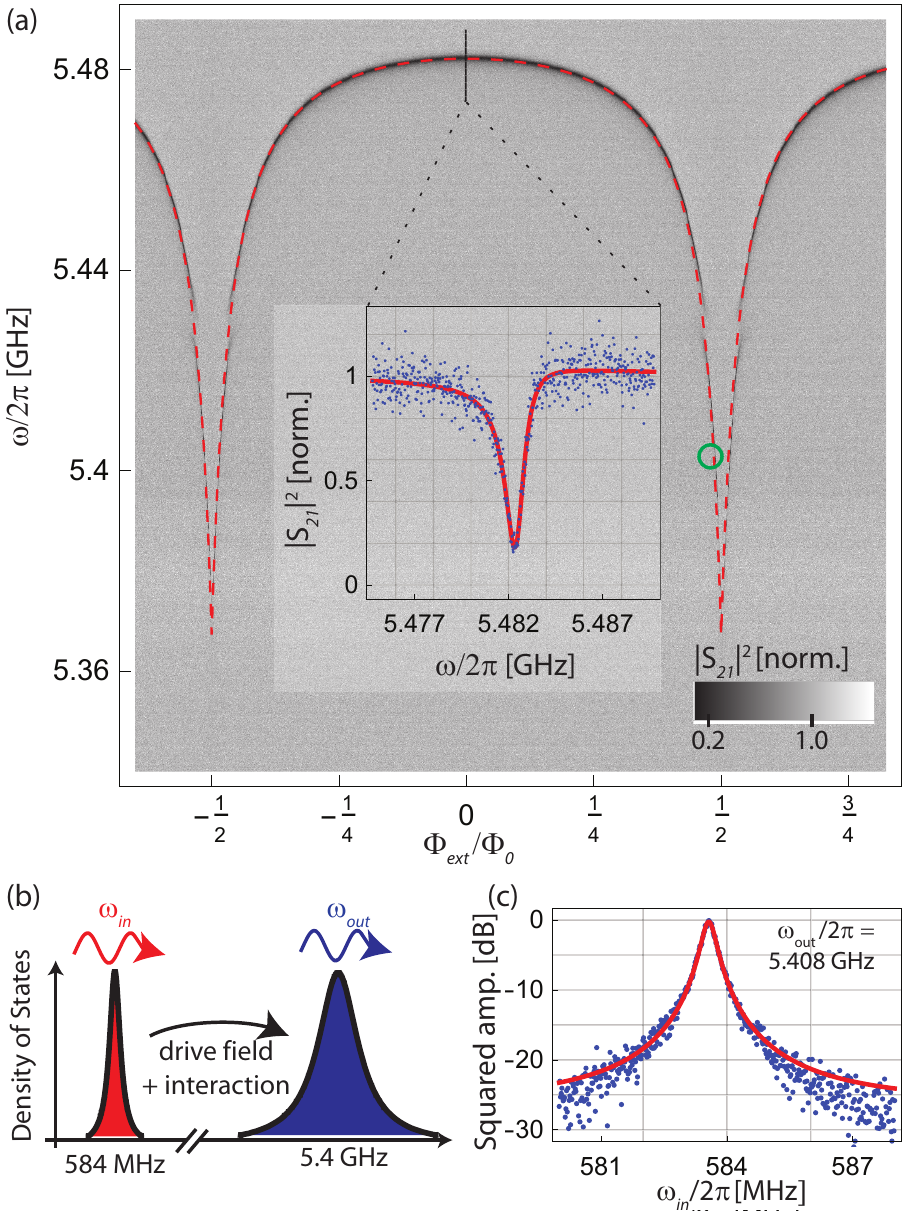}
\caption{(a) Measured transmission spectrum $S_{21}$ of the tunable high frequency resonator \emph{vs.} $\Phi_{\rm ext}$. The inset shows a line cut at around zero flux bias together with a fit to the theory. A model fit to the measured resonance frequencies as a function of magnetic flux is shown as the dashed red line. The green circle indicates the chosen bias point. (b) Schematic of the two-tone spectroscopy experiment. A drive field together with the nonlinear coupling mediates coherent upconversion from a low frequency input field $\omega_{\rm in}$ to the output field $\omega_{\rm out}$. (c) Measured squared amplitude of the upconverted field at frequency $\omega_{\rm out}$ as a function of $\omega_{\rm in}$ (blue points)  normalized to its maximum value and Lorentzian fit (solid red line).}
\label{fig:2}
\end{figure}

We first probe the resonance frequency $\omega_{a}$ of the high frequency resonator as a function of external magnetic flux $\Phi_{\rm ext}$ [see \figref{fig:2}(a)]. The flux is applied through a superconducting coil mounted below the sample holder. By fitting the individual transmission spectra (see inset for an example) we find a maximum resonance frequency of about $\omega_{a}/2\pi=5.48\,$GHz, an internal loss rate $\kappa_{\rm int}/2\pi \approx 0.5\,$MHz, and a external coupling rate to the transmission line $\kappa_{\rm ext}/2\pi \approx 0.7\,$MHz at $\Phi_{\rm ext}$. As expected from the increase in Josephson inductance, the resonance frequency decreases when tuning $\Phi_{\rm ext}$ towards half a flux quantum. The flux dependence of the measured resonance frequencies is well fit by a model (dashed red line) taking the finite loop inductance into account \cite{Pogorzalek2016}, which allows us to estimate the resonator geometric inductance $L_{\rm geo} \approx 20\,$nH and the self-inductance of the SQUID loop $L_{\rm loop} \approx 0.05\,$nH. The maximal Josephson inductance $L_{J,{\rm max}} \approx 0.11\,$nH entering the model is estimated based on the normal state resistance of identically fabricated SQUIDs measured at room temperature.

In order to increase the bare coupling strength $g_0$ to the low frequency resonator we choose $\Phi_{\rm ext}\approx\Phi_0/2$ [green circle in \figref{fig:2}(a)] at which the gradient $\partial \omega_{a}/ \partial \Phi_{\rm ext} \approx 1.7\,{\rm GHz}/\Phi_0$ becomes large resulting in an estimated $g_0/2\pi \approx 15\,$kHz.  We note that the choice $L_{\rm geo}/L_{J,{\rm max}}\gg1$ for our sample allows us to achieve a large gradient close to half a flux quantum while keeping the self-Kerr nonlinearity $\hbar K \approx E_c (L_J/L_{\rm tot})^3$ with $E_c=e^2\omega_{a}^2 L_{\rm tot}/2$ small. The self-Kerr nonlinearity imposes a limitation to the maximum applicable drive power and thus the achievable sideband induced coupling strength. For the chosen bias point we estimate a nonlinearity constant of $K/2\pi\approx 20\,$kHz, which is of similar order as typical values in parametric amplifiers \cite{Eichler2014} and about ten times larger than the residual nonlinearity reported from 3D cavity experiments \cite{Vlastakis2013}.

The nonlinear nature of the coupling Hamiltonian between the two modes allows one to enhance the bare coupling strength with an additional coherent drive field. By applying a drive field at the red sideband defined by the difference frequency between the two modes $\omega_{d}=\omega_{a}-\omega_{b}$, the coupling Hamiltonian in a rotating wave approximation takes the standard Jaynes-Cummings form
\begin{eqnarray}
H_{int}/\hbar \approx g ( \tilde{a}^\dagger b + \tilde{a} b^\dagger ),
\end{eqnarray}
where $g = g_0 \alpha_d$, and the field operator $\tilde{a}=a-\alpha_d$ describes fluctuations around the average coherent drive field  $\alpha_d=\langle a \rangle$ in the resonator \cite{Aspelmeyer2013}. The coherent field $\alpha_d=\sqrt{n_d}$ thus equals the square root of the number of coherent drive photons $n_d$ in the resonator and is thus dimensionless. The above resonance condition for the drive field can be understood intuitively from an energy conservation argument. A low frequency photon can be converted into a high frequency photon through the absorption of a photon from the drive field, while a high frequency photon can be converted into a low frequency one by emitting a photon into the drive field.
\begin{figure}[t!]
\includegraphics[scale=0.93]{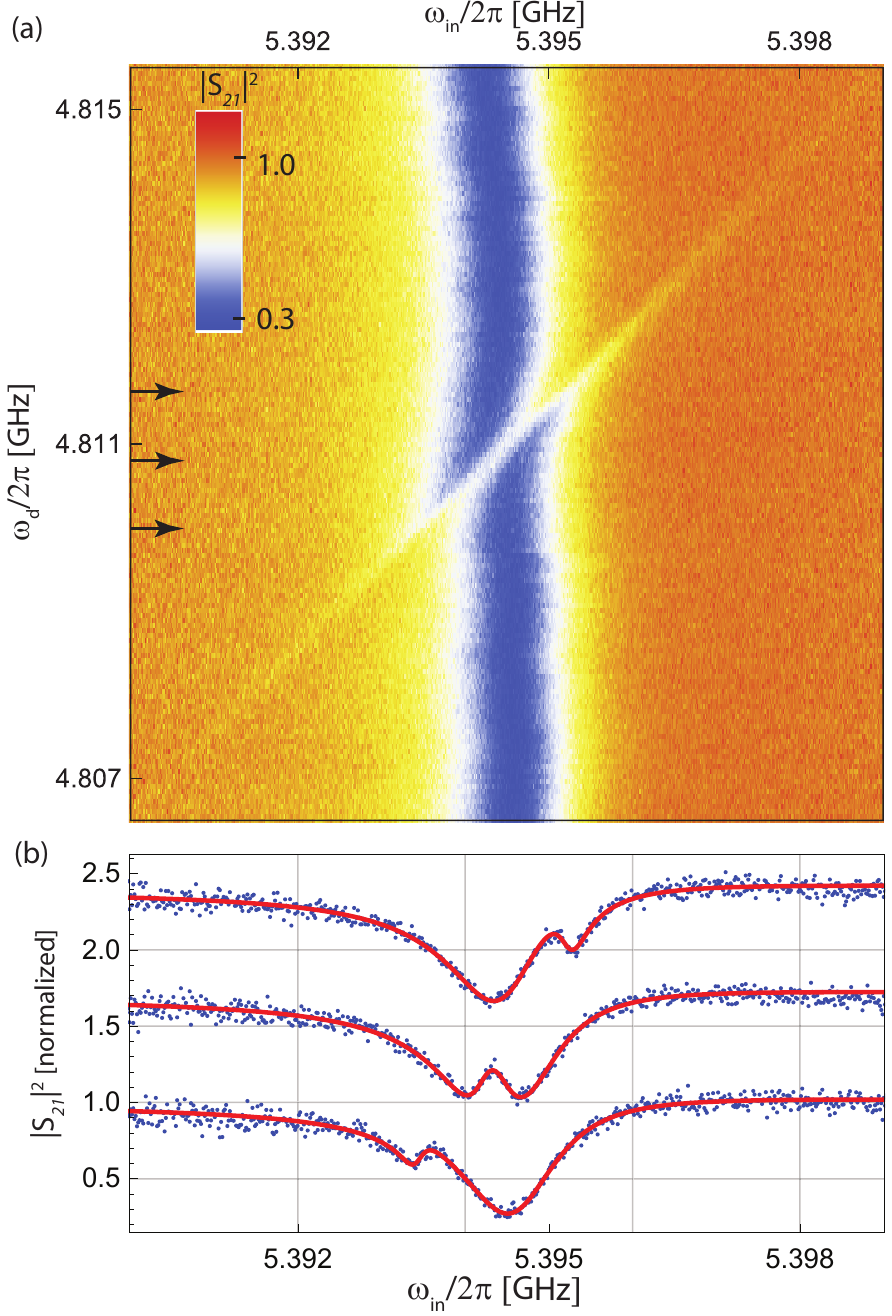}
\caption{(a) Transmission spectroscopy data $|S_{21}|^2$ as a function of probe frequency $\omega_{\rm in}$ for varying drive frequencies $\omega_{d}$.
(b) Individual transmission spectra at selected drive frequencies as indictated by the arrows in (a). The data (blue points) is fit to the input-output formula (red lines) in Eq.~(\ref{eq:S21}). Individual data sets are offset for clarity.}
\label{fig:3}
\end{figure}

We first use this drive induced coupling to probe the low frequency resonator in a two-tone spectroscopy experiment. Here, we apply both an input tone at variable frequency $\omega_{\rm in}$ to excite the low frequency resonator, and a drive field at frequency $\omega_{d} = \omega_{\rm out}-\omega_{\rm in}$, where the frequency $\omega_{\rm out}/2\pi = 5.408\,$GHz is kept constant at the frequency of the high frequency resonator. Both tones are applied through the feedline passing the high frequency resonator.
At the output we measure the amplitude of the radiation field at frequency $\omega_{\rm out}$ by employing standard analog down-conversion techniques and FPGA electronics. When the frequency of the input signal $\omega_{\rm in}$ is close to $\omega_{b}$, the nonlinear interaction mediates wave-mixing of the two fields resulting in a measurable upconverted signal at the sum frequency $\omega_{\rm out}=\omega_d + \omega_{\rm in}$ [see \figref{fig:2}(b)]. Indeed, we observe this coherent upconversion of the input signal in a 340$\,$kHz wide Lorentzian band around the resonance frequency $\omega_{b}/2\pi\approx583.5\,$MHz as shown in \figref{fig:2}(c).

In the above upconversion experiment we have chosen a moderate drive amplitude $\alpha_d \approx 9$, which results in an effective coupling of about $g/2\pi\approx120\,$kHz. We next study the sideband induced coupling in the limit of stronger drive fields $\alpha_d \approx 19$ by directly measuring the transmission spectrum $S_{21}$ as a function of probe frequency $\omega_{\rm in}$ and drive frequency $\omega_{d}$. When driving the system close to $\omega_d/2\pi = 4.811\,$GHz we observe an avoided crossing characteristic of resonant coupling of two modes. By fitting individual traces of the transmission spectrum to a model resulting from input-output theory  \cite{Gardiner1985}
\begin{eqnarray}
S_{21} &=& \frac{i {\kappa_{\rm int}} (\frac{i \gamma_{\rm}}{2} + \omega_{\rm in} -\omega_d - {\omega_{b}})}{g^2-(\frac{i\gamma_{\rm}}{2} + \omega_{\rm in} -\omega_d-{\omega_{b}}) (\frac{i\kappa}{2}+ \omega_{\rm in} -{\omega_{a}})}+e^{i \theta},
\nonumber
\\
\label{eq:S21}
\end{eqnarray}
we extract the intrinsic loss rate $\gamma_{\rm}/2\pi\approx300\,$kHz of the low frequency resonator, the linewidth of the high frequency resonator $\kappa/2\pi\equiv (\kappa_{\rm int} + \kappa_{\rm ext})/2\pi =1.5\,$MHz, and the effective coupling strength $g/2\pi\approx280\,$kHz for this particular drive field. The parameter $\theta/2\pi\approx-0.04$ accounts for the slight asymmetry of the tails of the resonance dip, which is a characteristic feature of resonators side-coupled to a feedline \cite{Khalil2012}. From the difference between $\omega_{a}$ and $\omega_{d}$ at which the coupling becomes resonant, we identify the frequency of the low frequency resonator $\omega_{b}/2\pi\approx583.53\,$MHz, which is in perfect agreement with the frequency we have found in the two-tone spectroscopy experiment and also by directly probing the transmission through the low frequency resonator in similar devices at a temperature of $\sim1.8\,$K. The fitted parameters correspond to a cooperativity of $4g^2/\kappa\gamma\approx 0.7$. In contrast to typical optomechanical systems, in which the cooperativity is often limited by small coupling strengths $g_0$, the cooperativity in our superconducting device is limited by the maximum applicable drive field and by the linewidth of the low frequency resonator. By increasing the quality factor and by further optimizing the circuit parameters in future devices such as the SQUID loop size and the mutual inductance between the two resonators, significant enhancements of the cooperativity seem feasible.
\begin{figure}[t!]
\includegraphics[scale=0.92]{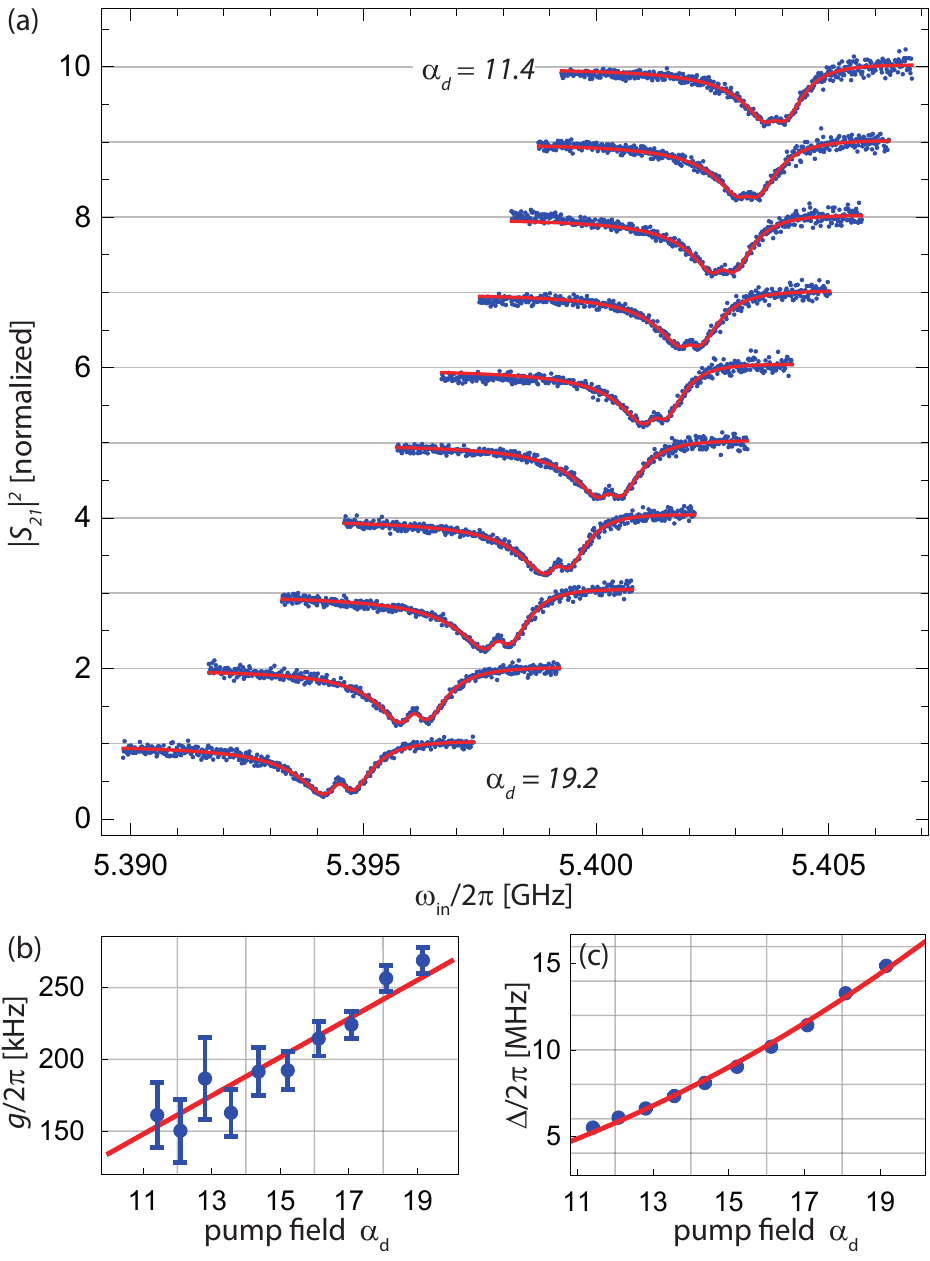}
\caption{(a) Data and fit to Eq.~(\ref{eq:S21}) of the measured transmission spectra on resonance for various drive amplitudes successively increasing by a factor $1.06$ from $\alpha_d=11.4$ to 19.2. The drive frequency $\omega_d$ is adjusted in each measurement to account for the power dependent Stark shift of the high frequency resonator. Individual data sets are offset by integer values for clarity. (b)-(c) Coupling strength and Stark shift {{vs.}}  $\alpha_d$ with linear and quadratic fits, respectively. Error bars are standard errors resulting from the fit of the data in (a) to Eq.~(\ref{eq:S21}).}
\label{fig:4}
\end{figure}

In order to characterize the observed coupling mechanism in more detail and to unambiguously show that the coupling is induced by the sideband drive, we measure transmission spectra comparable to the ones shown in \figref{fig:3} for varying drive power. The power dependent transmission spectra for which the coupling becomes resonant are plotted in \figref{fig:4}. We mainly observe two effects when increasing the drive power. First, the high frequency mode is shifted to lower frequencies. We attribute this behavior to a Stark shift $\Delta = 2 K |\alpha_d|^2$ proportional to the square of the drive field \cite{Ong2013}. Second, the coupling strength $g$ increases. By fitting the shown transmission spectra to Eq.~(\ref{eq:S21}) we extract the two paramters $\Delta$ and $g$ as shown in Figs.~\ref{fig:4}(b-c). While the coupling strength increases linearly with the drive field the Stark shift exhibits the expected quadratic dependence. The measured Stark shift together with the estimated nonlinearity allows us to determine the amplitude of the drive field $\alpha_d$ inside the resonator. Based on this absolute scaling of the sideband drive amplitude we can also estimate the bare coupling strength $g_0$ as the slope of the linear fit in \figref{fig:4}(b), which we find to be $g_0/2\pi\approx13\,$kHz and thus in reasonable agreement with the coupling strength we have calculated above based on independently estimated device parameters. As mentioned earlier, the Kerr nonlinearity of the high frequency resonator limits the maximum drive amplitude we can apply to the system. Indeed, we find that when further increasing the drive power the internal quality factor of the high frequency resonator decreases and eventually becomes unstable.

In conclusion, we demonstrated longitudinal coupling of two superconducting resonators detuned by more than three octaves, similar to what has been demonstrated with optomechanical devices. The nonlinear nature of the coupling mechanism allows us to employ coherent drive fields to control and enhance the effective coupling strength. By bridging the large energy gap between the two resonators, such a drive field mediates coherent frequency conversion between the two coupled modes. The SQUID based coupling scheme is very general and could also be used to couple qubits with resonators. Our experimental results suggest that entering a parameter regime in which the low frequency linewidth is dominated by Purcell decay into the high frequency resonator is feasible in the future. In this case, the low frequency resonator subject to finite thermal population could potentially be cooled into its quantum ground state, as has been demonstrated with optomechanical devices \cite{Teufel2011}. Another interesting direction could employ a blue sideband drive field to mediate an effective two-mode squeezing interaction. A combination of  blue and red sideband drive fields could be useful for amplification beyond the standard gain-bandwidth limit \cite{Metelmann2014}. Extending the device to multimode systems would enable the development of superconducting circulators and non-reciprocal devices \cite{Barzanjeh2017,Bernier2017,Rosenthal2017}, and for the creation of artificial gauge fields in cavity arrays \cite{Walter2016,Fang2017}.
%
%
\end{document}